\documentstyle[12pt]{article}

\begin{document}

\author{\baselineskip20pt{\normalsize {\bf Cheng Chen}} \\
\baselineskip20pt{\normalsize Institute of Plasma Physics, Academia Sinica, }%
\\
{\normalsize Hefei, Anhui, 230031, P.R. China}\\
{\normalsize and}\\
{\normalsize {\bf G. Cheng}}\\
{\normalsize CCAST (World Laboratory), P.O. Box 8730, Beijing 100080, P.R.
China}\\
{\normalsize Fundamental Physical Center, USTC., Hefei, Anhui, 230026, P.R.
China$\thanks{%
Mailing address}$}}
\title{Anomalous Dimension in the Solution of the Barenblatt's Equation }
\date{}
\maketitle

\begin{abstract}
\baselineskip20pt A new method is presented to obtain the anomalous
dimension in the solution of the Barenblatt's equation. The result is the
same as that in the renormalization group (RG) approach. It 
gives us insight on the perturbative solution of the
Barenblatt's equation in the RG approach. Based on this discussion, an
improvement is made to take into account, in more complete way,
the nonlinear effect, which is
included in the Heaviside function in higher orders. 
This improved result is better than that in RG approach.
\end{abstract}

\baselineskip20pt

\noindent {\large {\bf I.} {\bf Introduction}}

\ {\large \ }

\ {\normalsize Renormalization and the renormalization group (RG) method are 
powerful mathematical physics tool in quantum field theory and condensed
matter physics.$^{1,2}$ Recently it has been found that there are also
important applications in dynamical process, asymptotic analysis and
perturbation theory.$^{3-5}$ The fractal and chaos behaviors have been
studied with the RG approach$^3$. The introduction of the intermediate
asymptotic idea is very useful. It can help us to search for the connection
between the similarity, fractal and anomalous dimension of the RG theory$^3$%
. But there are still ambiguities in some aspects. For example, in quantum
field theory and in dynamical theory, it is well known that the RG approach
can improve the results of the perturbation calculation.$^{1-5}$ It can
remove the divergence and lead to higher leading order behavior from a lower
order perturbation calculation. It is also clear that the divergence and the
renormalization are physical requirements for some theories. But some of the
divergences occur without apparent physical reasons. 
One question involves the improvement that arise from the RG perturbative 
calculations.
In this paper we study this problem by analyzing a typical example. }

{\normalsize Barenblatt's equation has many good characteristics. It has
strong physical background and appears in many different disciplines.$^6$ It
has been studied by applying the method of similarity analysis. It can be
solved in terms of the parabolic cylinder functions in the intermediate
asymptotic region. Its perturbation solution is divergent and has anomalous
dimension in its asymptotic behavior. It is renormalizable and can be
treated with the renormalization and RG approach$^{3,4}$. But there are
still some ambiguities as mentioned above. }

{\normalsize In this paper we study improvements on the
perturbation solution of the Barenblatt's equation by renormalization and RG
treatment and, attempt to expose some general principles through the discussion
of this example. Based on the above analysis, an improved result is obtained
by considering the nonlinear effect, which is included in the Heaviside
function in higher orders, in more complete way. }

\ 

{\normalsize \noindent }{\large {\bf II. Anomalous dimension calculation by
the RG method}}

\ 

{\normalsize In this section, we repeat some results that are well known,
for consistency in notation and discussion style. }

{\normalsize We consider the Barenblatt's equation in one space
dimension 
\begin{equation}
\label{1}\partial _tu\left( x,t\right) -\frac 12\partial _x^2u\left(
x,t\right) =\frac \varepsilon 2\theta \left( -\partial _tu\left( x,t\right)
\right) \partial _x^2u\left( x,t\right) , 
\end{equation}
with symmetrical initial condition. It describes a nonlinear diffusive
process with non-conservative `mass'. }

{\normalsize Introduce a change of variables 
\begin{equation}
\label{2}\xi =\xi \left( x,t\right) =\frac x{\sqrt{t+l^2}}\qquad and\qquad
T=T\left( x,t\right) =t+l^2. 
\end{equation}
Then eq.(1) reads 
\begin{equation}
\label{3}\partial _Tu(\xi ,T)-\frac \xi {2T}\partial _\xi u(\xi ,T)-\frac
1{2T}\partial _\xi ^2u(\xi ,T)=\frac \varepsilon {2T}\theta \left(
(-\partial _T+\frac \xi {2T}\partial _\xi )u(\xi ,T)\right) \partial _\xi
^2u(\xi ,T). 
\end{equation}
We use an initial condition of $u(\xi ,T)\,$ as 
\begin{equation}
\label{4}u(\xi ,l^2)=\frac{m_0}{\sqrt{2\pi l^2}}\exp \left( -\frac{\xi ^2}%
2\right) . 
\end{equation}
It corresponds to the initial condition for $u(x,t)$ in the variables $x$
and $t$ as follows: 
\begin{equation}
\label{5}u(x,0)=\frac{m_0}{\sqrt{2\pi l^2}}\exp \left( -\frac{x^2}{2l^2}%
\right) . 
\end{equation}
For a general diffusion equation, there is a finite solution under this
initial condition. In this case, the initial condition is singular when $l^2$
approaches zero as a $\delta $-function. $m_0$ is a finite constant. But it
is not so in the case of Barenblatt's equation$^{3,4,6}$. It must be
renormalized to get a finite solution. The $m_0$ is no longer a constant
when the $l^2$ changes. It becomes infinite when the support of the initial
state approaches zero. We will discuss that in the end of this section. }

{\normalsize Introducing the expansion 
\begin{equation}
\label{6}u(\xi ,T)=u_0(\xi ,T)+\varepsilon u_1(\xi ,T)+\cdot \cdot \cdot
\cdot ,
\end{equation}
one has a sequence of equations: 
\begin{equation}
\label{7}2T\partial _Tu_0(\xi ,T)-\xi \partial _\xi u_0(\xi ,T)-\partial
_\xi ^2u_0(\xi ,T)=0,
\end{equation}
\begin{equation}
\label{8}2T\partial _Tu_1(\xi ,T)-\xi \partial _\xi u_1(\xi ,T)-\partial
_\xi ^2u_1(\xi ,T)=\theta \left( (-\partial _T+\frac \xi {2T}\partial _\xi
)u_0(\xi ,T)\right) \partial _\xi ^2u_0(\xi ,T),
\end{equation}
and so on.{\normalsize \ }

{\normalsize Furtherly, we assume that the initial condition of the $u_0(\xi
,T)$ is: 
\begin{equation}
\label{9}u_0(\xi ,l^2)=\frac{m_0}{\sqrt{2\pi l^2}}\exp \left( -\frac{\xi ^2}%
2\right) . 
\end{equation}
Then the initial condition of the $u_n(\xi ,T)$ is: 
\begin{equation}
\label{10}u_n(\xi ,l^2)=0\qquad n\geq 1. 
\end{equation}
The solution of the zero-order approximation is: 
\begin{equation}
\label{11}u_0\left( \xi ,T\right) =\frac{m_0}{\sqrt{2\pi T}}\exp \left( -%
\frac{\xi ^2}2\right) . 
\end{equation}
Since one has 
\begin{equation}
\label{12}\left( -\partial _T+\frac \xi {2T}\partial _\xi \right) u_0(\xi
,T)=\left( \frac 1{2T}-\frac{\xi ^2}{2T}\right) u_0(\xi ,T), 
\end{equation}
one obtains 
\begin{equation}
\label{13}\theta \left( (-\partial _T+\frac \xi {2T}\partial _\xi )u_0(\xi
,T)\right) =\theta \left( 1-\left| \xi ^2\right| \right) . 
\end{equation}
This is the Heaviside function of the Barenblatt's equation in its lowest
order. We can write the first-order equation as: 
\begin{equation}
\label{14}2T\partial _Tu_1(\xi ,T)-\xi \partial _\xi u_1(\xi ,T)-\partial
_\xi ^2u_1(\xi ,T)=\theta \left( 1-\left| \xi ^2\right| \right) \partial
_\xi ^2u_0(\xi ,T). 
\end{equation}
From the similarity analysis we know that the solution of the Barenblatt's
equation can be written in a form of variable separation: 
\begin{equation}
\label{15}u(\xi ,T)=F\left( T\right) G(\xi ). 
\end{equation}
It is adequate when we are only interested in the behavior of the solution
in the intermediate asymptotic region. }

{\normalsize Expanding 
\begin{equation}
\label{16}F\left( T\right) =F_0\left( T\right) +\varepsilon F_1\left(
T\right) +\cdot \cdot \cdot \cdot , 
\end{equation}
and 
\begin{equation}
\label{17}G(\xi )=G_0(\xi )+\varepsilon G_1(\xi )+\cdot \cdot \cdot \cdot , 
\end{equation}
we have 
\begin{equation}
\label{18}u_0(\xi ,T)=F_0\left( T\right) G_0(\xi ) 
\end{equation}
and 
\begin{equation}
\label{19}u_1(\xi ,T)=F_0\left( T\right) G_1(\xi )+F_1\left( T\right)
G_0(\xi ). 
\end{equation}
We now take 
\begin{equation}
\label{20}F_0\left( T\right) =\frac{m_0}{\sqrt{2\pi T}}\quad and\quad
G_0(\xi )=\exp \left( -\frac{\xi ^2}2\right) . 
\end{equation}
The solution in the first-order approximation is 
\begin{equation}
\label{21}u^{(1)}(\xi ,T)=F_0\left( T\right) G_0(\xi )+\varepsilon \left(
F_0\left( T\right) G_1(\xi )+F_1\left( T\right) G_0(\xi )\right) . 
\end{equation}
It tells us that the approximation solution not can be expressed in the form
of variable separation in finite order of the perturbative theory. But we
will see that the RG solution is in this form. Thus, we shall pursue along
this line and accept a corresponding approximation, so that we can use the
variable separation solution. }

{\normalsize Inserting eq.$\left( 19\right) $ into eq.$\left( 14\right) $
yields 
\begin{equation}
\label{22}2T\left( F_1^{^{\prime }}G_0+F_0^{^{\prime }}G_1\right) -\xi
\left( F_1G_0^{^{\prime }}+F_0G_1^{^{\prime }}\right) -\left(
F_1G_0^{^{\prime \prime }}+F_0G_1^{^{\prime \prime }}\right) =\theta \left(
1-\left| \xi \right| \right) F_0G_0^{^{\prime \prime }}, 
\end{equation}
where 
\begin{equation}
\label{23}F_i^{^{\prime }}=\frac d{dT}F_i\left( T\right) ,\quad
G_i^{^{\prime }}=\frac d{d\xi }G_i(\xi ), \qquad and \qquad G_i^{^{\prime 
\prime }}=\frac{d^2}{d\xi ^2}G_i(\xi ),\qquad i=0,1. 
\end{equation}
}

{\normalsize Now, we search for the solution of the $F_1(T)$. Integrate this
equation with respect to the $\xi $ from $-\infty $ to $+\infty $ and let 
\begin{equation}
\label{24}\int\limits_{-\infty }^\infty G_1(\xi )d\xi =a.
\end{equation}
After simple integral calculation and the application of the conditions 
\begin{equation}
\label{25}\lim \limits_{\xi \rightarrow \pm \infty }\xi G_1(\xi )=0\qquad
and\qquad \lim \limits_{\xi \rightarrow \pm \infty }G_1^{^{\prime }}(\xi )=0,
\end{equation}
one obtains 
\begin{equation}
\label{26}2aF_1(T)TF_0^{^{\prime }}(T)+2\sqrt{2\pi }TF_1^{^{\prime
}}(T)+aF_0(T)+\sqrt{2\pi }F_1(T)=-\frac 2{\sqrt{e}}F_0(T).
\end{equation}
Substituting $F_0\left( T\right) =m_0/\sqrt{2\pi T}$ and $F_0^{^{\prime
}}\left( T\right) =-F_0\left( T\right) /2T$ into eq.$\left( 26\right) $, we
get 
\begin{equation}
\label{27}TF_1^{^{\prime }}(T)+\frac 12F_1(T)+\frac 1{\sqrt{2\pi e}}F_0(T)=0.
\end{equation}
The solution of eq.$(27)$ is 
\begin{equation}
\label{28}F_1(T)=-F_0(T)\beta \ln \frac T{l^2},
\end{equation}
where 
\begin{equation}
\label{29}\beta =\frac 1{\sqrt{2\pi e}}.
\end{equation}
For obtaining this solution, we have made use of the initial condition 
\begin{equation}
\label{30}F_n(l^2)=0\qquad n\geq 1.
\end{equation}
The first-order approximation of $F(T)$ is 
\begin{equation}
\label{31}F^{\left( 1\right) }(T)=F_0(T)\left( 1-\alpha \ln \frac
T{l^2}\right) ,
\end{equation}
where 
\begin{equation}
\label{32}\alpha =\varepsilon \beta =\frac \varepsilon {\sqrt{2\pi e}}.
\end{equation}
It is the anomalous dimension and is well known in the solution of the
Barenblatt's equation in the intermediate asymptotic region by application
of RG approach$^{3,4}$. }

{\normalsize In the following discussion, we keep the $G(\xi )$ in its
zero-order approximation without considering the higher-order corrections.
That is to put $G_n(\xi )=0,\,n\geq 1.$ Then the solution of the first order
is 
\begin{equation}
\label{33}u^{(1)}(\xi ,T)=F_0(T)G_0(\xi )\left( 1-\alpha \ln \frac
T{l^2}\right) . 
\end{equation}
It has been proved that this approximation does not have any effect on the
function $F^{\left( 1\right) }(T)$ when we make the RG treatment. We will
discuss the space behavior in a separate paper. }

{\normalsize For getting a finite but non-zero asymptotic solution, the
`mass' 
\begin{equation}
\label{34}m_0=\int\limits_{-\infty }^\infty dxu\left( x,t\right) 
\end{equation}
not can remain a constant. It is necessary to approach infinity when the
support of the initial function approaches zero. To obtain a finite solution
of Barenblatt's equation, the $u\left( x,0\right) $ must be more singular
than the $\delta $-function. It is different from the diffusive equation in
regular cases. Moreover, the values of the non-zero order terms are not
small, but instead, very large in the intermediate asymptotic region. The
renormalization is required. It is well known that the renormalization and
RG approach not only remove the divergence but also improve the perturbation
expansion solution to higher leading order.$^{1-5}$ }

{\normalsize Let 
\begin{equation}
\label{35}m_0=m_R(\mu )Z(\mu ,l^2)
\end{equation}
and 
\begin{equation}
\label{36}Z(\mu ,l^2)=1+\alpha \ln \frac \mu {l^2}.
\end{equation}
From the work of Goldenfeld et al,$^{3,4}$ we know that the result of the
renormalization and RG treatment is: 
\begin{equation}
\label{37}u_R^{(1)}(\xi ,T)=\frac{m_R(l^2)}{\sqrt{2\pi T}}\left( \frac{l^2}%
T\right) ^\alpha \exp \left( -\frac{\xi ^2}2\right) .
\end{equation}
It is easy to check that the initial condition, the boundary conditions and
the connection condition are all satisfied when the $l^2$ is taken in
non-zero value. But we must remember that the renormalization mass will
approach the infinite when $l^2\rightarrow 0,$ in order to keep the solution
finite. Usually we require 
\begin{equation}
\label{38}m_R(l^2)\times l^{2\alpha }=finite.
\end{equation}
There is an explanation for the improvement of the RG approach.$^7$ But the
whole case is not very clear yet. We will discuss some aspects of it in
following sections.}

\ 

{\normalsize \noindent }{\large {\bf III.} {\bf Anomalous dimension
calculation by summing }$F\left( T\right) ${\bf \ to the infinite order}}

\ {\large \ }

{\normalsize Now we discuss the improvement to the 
perturbative solution of the Barenblatt's equation in the RG approach. 
We do this by using the
approximation suggested in the previous section. We are
especially interested in the behavior of the intermediate asymptotic region, 
in which $%
\xi $ and $t$ are far from the zero point but not in infinite. We shall
prove that the result of the renormalization and RG approach is equivalent
to the infinite sum of the series of $F(T).$ }

{\normalsize We assume the solution of eq.$\left( 3\right) $ in a form of
variable separation in any order. We have known that it is not 
a suitable form
when the approximation order $n$ is finite. But, the higher the
approximation order is taken, the better the proposed form is. As the first
step, we keep the dependence on space in its lowest-order. Thus we set 
\begin{equation}
\label{39}u_n(\xi ,T)=f_n(T)u_0(\xi ,T). 
\end{equation}
At the same time, we also neglect a part of the nonlinear effects. We keep
the Heaviside function in its lowest order as $\theta \left( 1-\left| \xi
^2\right| \right) .$ Therefore we have the form of perturbation expansion of
the Barenblatt's equation. 
$$
2T\partial _Tu_n(\xi ,T)-\xi \partial _\xi u_n(\xi ,T)-\partial _\xi
^2u_n(\xi ,T)= 
$$
\begin{equation}
\label{40}=\theta \left( (-\partial _T+\frac \xi {2T}\partial _\xi )u_0(\xi
,T)\right) \partial _\xi ^2u_{n-1}(\xi ,T)\qquad \qquad n\geq 1. 
\end{equation}
We will prove that the RG anomalous dimension is just the result of summing
up the $F(T)$ series to infinite order in the above approximation. }

{\normalsize Inserting eqs.$\left( 13\right) $ and $\left( 39\right) $ into
eq.$\left( 40\right) $ yields 
\begin{equation}
\label{41}2Tu_0(\xi ,T)\frac d{dT}f_n(T)=\theta \left( 1-\left| \xi
^2\right| \right) f_{n-1}(T)\partial _\xi ^2u_0(\xi ,T)\qquad \qquad n\geq
1. 
\end{equation}
The initial condition 
\begin{equation}
\label{42}u_n(\xi ,l^2)=0\qquad \qquad n\geq 1 
\end{equation}
becomes 
\begin{equation}
\label{43}f_n(l^2)=0\qquad \qquad n\geq 1. 
\end{equation}
}

{\normalsize When $n=1$, one has 
\begin{equation}
\label{44}2Tu_0(\xi ,T)\frac d{dT}f_1(T)=\theta \left( 1-\left| \xi
^2\right| \right) \partial _\xi ^2u_0(\xi ,T).
\end{equation}
Integrating it with respect to $\xi $ from $-\infty $ to $+\infty $ gives 
\begin{equation}
\label{45}T\frac d{dT}f_1(T)=-\frac 1{\sqrt{2\pi e}}=-\beta .
\end{equation}
Using  $f_n(l^2)=0,$ }$n\geq 1,$ {\normalsize the solution of eq.$\left(
45\right) $ is 
\begin{equation}
\label{46}f_1(T)=-\frac 1{\sqrt{2\pi e}}\ln \frac T{l^2}=\ln \left( \frac
T{l^2}\right) ^{-\beta }.
\end{equation}
}

{\normalsize When $n=2,$ one has 
\begin{equation}
\label{47}2Tu_0(\xi ,T)\frac d{dT}f_2(T)=\theta \left( 1-\left| \xi
^2\right| \right) f_1(T)\partial _\xi ^2u_0(\xi ,T).
\end{equation}
Doing the same integration as we have done to the eq.$(44),$ we obtain 
\begin{equation}
\label{48}2T\frac{m_0}{\sqrt{T}}\frac
d{dT}f_2(T)=f_1(T)\int\limits_{-1}^1d\xi \partial _\xi ^2u_0(\xi ,T).
\end{equation}
Due to the identity 
\begin{equation}
\label{49}2T\frac{m_0}{\sqrt{T}}\frac d{dT}f_1(T)=\int\limits_{-1}^1d\xi
\partial _\xi ^2u_0(\xi ,T),
\end{equation}
we have 
\begin{equation}
\label{50}\frac d{dT}f_2(T)=f_1(T)\frac d{dT}f_1(T).
\end{equation}
Making use of  $f_n(l^2)=0,$ }$n\geq 1,$ {\normalsize we obtain 
\begin{equation}
\label{51}f_2(T)=\frac 12\left[ f_1(T)\right] ^2.
\end{equation}
Using the principle of mathematical induction, we now prove that 
\begin{equation}
\label{52}f_m(T)=\frac 1{m!}\left[ f_1(T)\right] ^m.
\end{equation}
Suppose that eq$.(52)$ is true for $m=n-1,$ that is to say, 
\begin{equation}
\label{53}f_{n-1}(T)=\frac 1{\left( n-1\right) !}\left[ f_1(T)\right] ^{n-1}.
\end{equation}
We are going to prove that it is true for $m=n.$ Integrating eq.$\left(
41\right) $ with respect to $\xi $ from $-\infty $ to $+\infty $ gives 
\begin{equation}
\label{54}2T\frac{m_0}{\sqrt{T}}\frac
d{dT}f_n(T)=f_{n-1}(T)\int\limits_{-1}^1d\xi \partial _\xi ^2u_0(\xi ,T).
\end{equation}
Due to eq.$\left( 49\right) $, one has 
\begin{equation}
\label{55}\frac d{dT}f_n(T)=f_{n-1}(T)\frac d{dT}f_1(T).
\end{equation}
Substituting eq.$\left( 53\right) $ into it gives 
\begin{equation}
\label{56}\frac d{dT}f_n(T)=\frac 1{\left( n-1\right) !}\left[ f_1(T)\right]
^{n-1}\frac d{dT}f_1(T).
\end{equation}
Because $f_n(l^2)=0,$ the solution of eq.$\left( 56\right) $ is 
\begin{equation}
\label{57}f_n(T)=\frac 1{n!}\left[ f_1(T)\right] ^n.
\end{equation}
It is true for $m=1$ and $m=2.$ Thus we have proved the proposition for any $%
m$. }

{\normalsize In the infinite limit, we have 
\begin{equation}
\label{58}u\left( \xi ,T\right) =\left[ 1+\sum\limits_{n=1}^\infty \frac{%
\varepsilon ^n}{n!}\left[ f_1(T)\right] ^n\right] u_0\left( \xi ,T\right) . 
\end{equation}
Inserting eq.$\left( 46\right) $ into it yields 
\begin{equation}
\label{59}u\left( \xi ,T\right) =\left[ 1+\sum\limits_{n=1}^\infty \frac
1{n!}\left[ -\varepsilon \beta \ln \frac T{l^2}\right] ^n\right] u_0\left(
\xi ,T\right) , 
\end{equation}
that is 
\begin{equation}
\label{60}u\left( \xi ,T\right) =\exp \left[ -\varepsilon \beta \ln \frac
T{l^2}\right] u_0\left( \xi ,T\right) =\left[ \frac T{l^2}\right] ^{-\alpha
}u_0\left( \xi ,T\right) , 
\end{equation}
where 
\begin{equation}
\label{61}\alpha =\varepsilon \beta =\frac \varepsilon {\sqrt{2\pi e}}, 
\end{equation}
which is just the anomalous dimension given by the renormalization and RG
approach.}

\ 

{\normalsize \noindent }{\large {\bf IV.} {\bf An improvement on the
calculation of the anomalous dimension}}

\ 

{\normalsize In the previous section, we have kept the Heaviside function
and the function $G(\xi )$ in their lowest order in our calculation of the $%
F(T)$ to higher orders. It has led a partial loss of the nonlinear effect
which is included in the Heaviside function. Now we show that the
calculation of the anomalous dimension can be improved by the infinite
summing of the function $F(T)$ without the approximation of keeping the
Heaviside function in its lowest order to take account of a part of the lost
nonlinear effect. Now we still keep the solution in the form of variable
separation as eq.$(15),$ but only let the function $G(\xi )$ in it lowest
order. We shall use a trick that makes an analytic treatment possible. }

{\normalsize Take the approximation solution obtained in the previous
section 
\begin{equation}
\label{62}u\left( \xi ,T\right) =\left( \frac T{l^2}\right) ^{-\alpha
}u_0\left( \xi ,T\right) .
\end{equation}
as an initial solution. After inserting it into the Heaviside function, we
solve the $u_n(\xi ,T)$ and search for its infinite sum. Thus we obtain the
first iterative result. }

{\normalsize We put the first iterative solution into the Heaviside function
to solve the equation secondly. }

{\normalsize In this way, we search for the limit of the iterative solution
to infinite. If the limit exists, we can obtain the solution. }

{\normalsize An important observation is that the form of the solution
remains unchanged in the iterative process. 
\begin{equation}
\label{63}u^{\left( k\right) }\left( \xi ,T\right) =\left( \frac
T{l^2}\right) ^{-\alpha ^{\left( k\right) }}u_0\left( \xi ,T\right) . 
\end{equation}
The only change is the value of the anomalous dimension $\alpha ^{\left(
k\right) }.$ It simplifies the analysis considerably. }

{\normalsize We rewrite $\alpha $ as $\alpha ^{\left( 0\right) }$ 
\begin{equation}
\label{64}\alpha =\varepsilon \beta =\frac \varepsilon {\sqrt{2\pi e}%
}\rightarrow \alpha ^{\left( 0\right) }=\varepsilon \beta ^{\left( 0\right)
}. 
\end{equation}
The $\alpha ^{\left( k\right) }=\varepsilon \beta ^{\left( k\right) }$
represents the results after the $k$th iteration. }

{\normalsize We still need to solve the problem by summing
up the infinite expansion series, in each step of the iterative process. }

{\normalsize Putting eq.$\left( 63\right) $ into the Heaviside function of
the $n$th-order equation, we obtain the $n$th order and $(k+1)$th iteration
equation }

{\normalsize 
$$
2Tu_0(\xi ,T)\frac d{dT}f_n^{\left( k+1\right) }(T)= 
$$
\begin{equation}
\label{65}\theta \left( \left( -\partial _T+\frac \xi {2T}\partial _\xi
\right) \left( \left( \frac T{l^2}\right) ^{-\alpha ^{\left( k\right)
}}u_0(\xi ,T)\right) \right) f_{n-1}^{\left( k+1\right) }(T)\partial _\xi
^2u_0(\xi ,T). 
\end{equation}
where the super-index $\left( k\right) $ represents the result of the $k$th
iteration and the sub-index $n$ represents the $n$th order in the
perturbation expansion. The initial condition remains in the form of eq.$%
\left( 43\right) $. }

{\normalsize When $k=1$ and $n=1,$ one has 
\begin{equation}
\label{66}2Tu_0(\xi ,T)\frac d{dT}f_1^{\left( 1\right) }(T)=\theta \left(
\left( -\partial _T+\frac \xi {2T}\partial _\xi \right) \left( \left( \frac
T{l^2}\right) ^{-\alpha ^{\left( 0\right) }}u_0(\xi ,T)\right) \right)
\partial _\xi ^2u_0(\xi ,T), 
\end{equation}
where $f_0^{\left( k\right) }(T)=1$ has been used. Because of the identity 
\begin{equation}
\label{67}\left( -\partial _T+\frac \xi {2T}\partial _\xi \right) \left(
\left( \frac T{l^2}\right) ^{-\alpha ^{\left( 0\right) }}u_0(\xi ,T)\right)
=\left[ 2\alpha ^{\left( 0\right) }-\left( \xi ^2-1\right) \right] \frac
1{2T}\left( \frac T{l^2}\right) ^{-\alpha ^{\left( 0\right) }}u_0(\xi ,T), 
\end{equation}
one has 
\begin{equation}
\label{68}\theta \left( \left( -\partial _T+\frac \xi {2T}\partial _\xi
\right) \left( \left( \frac T{l^2}\right) ^{-\alpha ^{\left( 0\right)
}}u_0(\xi ,T)\right) \right) =\theta \left( 2\alpha ^{\left( 0\right)
}+1-\xi ^2\right) . 
\end{equation}
Substituting it into eq.$\left( 66\right) $ and integrating it with respect
to $\xi $ from $-\infty $ to $+\infty $ yields 
$$
2T\frac{m_0}{\sqrt{T}}\frac d{dT}f_1^{\left( 1\right)
}(T)=\int\limits_{-\infty }^\infty \theta \left( 2\alpha ^{\left( 0\right)
}+1-\xi ^2\right) \partial _\xi ^2u_0(\xi ,T) 
$$
\begin{equation}
\label{69}=\frac{m_0}{\sqrt{2\pi T}}\left[ -2\sqrt{\left( 1+2\alpha ^{\left(
0\right) }\right) }\exp \left( -\frac{1+2\alpha ^{\left( 0\right) }}2\right)
\right] , 
\end{equation}
that is, 
\begin{equation}
\label{70}\frac d{d\ln T}f_1^{\left( 1\right) }(T)=\frac 1{\sqrt{2\pi }%
}\left[ -\sqrt{\left( 1+2\alpha ^{\left( 0\right) }\right) }\exp \left( -%
\frac{1+2\alpha ^{\left( 0\right) }}2\right) \right] . 
\end{equation}
Using $f_1^{\left( 1\right) }(l^2)=0,$ we can obtain the solution of eq.$%
\left( 70\right) $%
\begin{equation}
\label{71}f_1^{\left( 1\right) }(T)=\frac{-\sqrt{\left( 1+2\alpha ^{\left(
0\right) }\right) }}{\sqrt{2\pi e^{\left( 1+2\alpha ^{\left( 0\right)
}\right) }}}\ln \frac T{l^2}=-\beta ^{\left( 1\right) }\ln \frac T{l^2}, 
\end{equation}
where 
\begin{equation}
\label{72}\beta ^{\left( 1\right) }=\frac{\sqrt{\left( 1+2\alpha ^{\left(
0\right) }\right) }}{\sqrt{2\pi e^{\left( 1+2\alpha ^{\left( 0\right)
}\right) }}}. 
\end{equation}
}

{\normalsize When $k=1$ and $n=2$, one has 
\begin{equation}
\label{73}2Tu_0(\xi ,T)\frac d{dT}f_2^{\left( 1\right) }(T)=\theta \left(
2\alpha ^{\left( 0\right) }+1-\xi ^2\right) f_1^{\left( 1\right)
}(T)\partial _\xi ^2u_0(\xi ,T).
\end{equation}
Integrating it with respect to $\xi $ from $-\infty $ to $+\infty $ yields 
\begin{equation}
\label{74}\frac d{dT}f_2^{\left( 1\right) }(T)=f_1^{\left( 1\right)
}(T)\frac d{dT}f_1^{\left( 1\right) }(T).
\end{equation}
Making use of  $f_n^{\left( 1\right) }(l^2)=0,$ }$n\geq 2,$ {\normalsize the
solution of eq.$(74)$ is 
\begin{equation}
\label{75}f_2^{\left( 1\right) }(T)=\frac 12\left[ f_1^{\left( 1\right)
}(T)\right] ^2.
\end{equation}
Using the principle of mathematical induction, it is easy to prove that 
\begin{equation}
\label{76}f_m^{\left( 1\right) }(T)=\frac 1{m!}\left[ f_1^{\left( 1\right)
}(T)\right] ^m.
\end{equation}
Suppose it is true for $m=n-1,$ that is to say 
\begin{equation}
\label{77}f_{n-1}^{\left( 1\right) }(T)=\frac 1{\left( n-1\right) !}\left[
f_1^{\left( 1\right) }(T)\right] ^{n-1}.
\end{equation}
We are going to prove that it is true for $m=n.$ We know that 
\begin{equation}
\label{78}2Tu_0(\xi ,T)\frac d{dT}f_n^{\left( 1\right) }(T)=\theta \left(
2\alpha ^{\left( 0\right) }+1-\xi ^2\right) f_{n-1}^{\left( 1\right)
}(T)\partial _\xi ^2u_0(\xi ,T).
\end{equation}
Integrating it with respect to $\xi $ from $-\infty $ to $+\infty $ gives 
\begin{equation}
\label{79}\frac d{dT}f_n^{\left( 1\right) }(T)=f_{n-1}^{\left( 1\right)
}(T)\frac d{dT}f_1^{\left( 1\right) }(T).
\end{equation}
Substituting eq.$\left( 77\right) $ into it yields 
\begin{equation}
\label{80}\frac d{dT}f_n^{\left( 1\right) }(T)=\frac 1{\left( n-1\right)
!}\left[ f_1^{\left( 1\right) }(T)\right] ^{n-1}\frac d{dT}f_1^{\left(
1\right) }(T).
\end{equation}
On account of $f_n^{\left( 1\right) }(l^2)=0,$ one can solve eq.$\left(
80\right) $ and obtain 
\begin{equation}
\label{81}f_n^{\left( 1\right) }(T)=\frac 1{n!}\left[ f_1^{\left( 1\right)
}(T)\right] ^n.
\end{equation}
We have proved that it is true for $m=1$ and $m=2$ . Thus the proposition is
true for any $m$. }

{\normalsize Now we sum the infinite series 
\begin{equation}
\label{82}u^{\left( 1\right) }\left( \xi ,T\right) =\left[
1+\sum\limits_{n=1}^\infty \frac{\varepsilon ^n}{n!}\left[ f_1^{\left(
1\right) }(T)\right] ^n\right] u_0\left( \xi ,T\right) . 
\end{equation}
Inserting the expression of $f_1^{\left( 1\right) }(T)$ into it, we obtain 
\begin{equation}
\label{83}u^{\left( 1\right) }\left( \xi ,T\right) =\left[
1+\sum\limits_{n=1}^\infty \frac 1{n!}\left[ -\varepsilon \beta ^{\left(
1\right) }\ln \frac T{l^2}\right] ^n\right] u_0\left( \xi ,T\right) . 
\end{equation}
That is 
\begin{equation}
\label{84}u^{\left( 1\right) }\left( \xi ,T\right) =\left[ \frac
T{l^2}\right] ^{-\alpha ^{\left( 1\right) }}u_0\left( \xi ,T\right) , 
\end{equation}
where 
\begin{equation}
\label{85}\alpha ^{\left( 1\right) }=\varepsilon \beta ^{\left( 1\right)
}\qquad and\qquad \beta ^{\left( 1\right) }=\frac{\sqrt{\left( 1+2\alpha
^{\left( 0\right) }\right) }}{\sqrt{2\pi e^{\left( 1+2\alpha ^{\left(
0\right) }\right) }}}. 
\end{equation}
The $\alpha ^{\left( 1\right) }$ is the anomalous dimension corrected to the
first-order of the $F(T).$ Now, we have finished the first iterative
calculation. }

{\normalsize When we make the $k$th iteration, the process is the same as
the $\left( k-1\right) $th one. The only difference is that we will use $%
\beta ^{\left( k\right) }$ in place of the $\beta ^{\left( k-1\right) }.$
Then it is easy to prove that 
\begin{equation}
\label{86}u^{\left( k\right) }\left( \xi ,T\right) =\left[ \frac
T{l^2}\right] ^{-\alpha ^{\left( k\right) }}u_0\left( \xi ,T\right) ,
\end{equation}
where 
\begin{equation}
\label{87}\alpha ^{\left( k\right) }=\varepsilon \beta ^{\left( k\right)
}\qquad and\qquad \beta ^{\left( k\right) }=\frac{\sqrt{\left( 1+2\alpha
^{\left( k-1\right) }\right) }}{\sqrt{2\pi e^{\left( 1+2\alpha ^{\left(
k-1\right) }\right) }}}.
\end{equation}
The $\alpha ^{\left( k\right) }$ is the anomalous dimension corrected to the 
$k$th order of the $F(T).$ }

{\normalsize If the iterative process is convergent, the limit exists. Let $%
\lim _{k\rightarrow \infty }\beta ^{\left( k\right) }=\beta ^{\left( \infty
\right) },$ which leads to   
\begin{equation}
\label{88}\beta ^{\left( \infty \right) }=\frac{\sqrt{\left( 1+2\varepsilon
\beta ^{\left( \infty \right) }\right) }}{\sqrt{2\pi e^{\left(
1+2\varepsilon \beta ^{\left( \infty \right) }\right) }}}.
\end{equation}
Then one has 
\begin{equation}
\label{89}u^{\left( \infty \right) }\left( \xi ,T\right) =\left[ \frac
T{l^2}\right] ^{-\alpha ^{\left( \infty \right) }}u_0\left( \xi ,T\right) ,
\end{equation}
where  
\begin{equation}
\label{90}{\normalsize \alpha ^{\left( \infty \right) }=\varepsilon \beta
^{\left( \infty \right) }}\ .
\end{equation}
and}$\qquad $%
\begin{equation}
\label{91}\alpha ^{\left( \infty \right) }=\frac{\varepsilon \sqrt{\left(
1+2\alpha ^{\left( \infty \right) }\right) }}{\sqrt{2\pi e^{\left( 1+2\alpha
^{\left( \infty \right) }\right) }}}
\end{equation}

{\normalsize \ It is obvious that eq.$\left( 91\right) $ has finite
solution, which is the anomalous dimension corrected by considering the
nonlinear effect of $F(T)$ to infinite order. }

{\normalsize The justification of the iterative method and its initial value
is that the solution is convergent and satisfies the Barenblatt's equation
in our approximation. }

{\normalsize We can compare the improved result with the numerical one given
by Barenblatt and the result obtained by RG approach as shown in Fig.1.
The improved result $\alpha ^{\left( \infty \right) }$ is better than $%
\alpha ^{\left( 0\right) }$. }

\ 

{\normalsize \noindent }{\large {\bf V.} {\bf Conclusion}}

\ 

{\normalsize We have calculated the anomalous dimension by summing the $F(T)$
to infinite order. When we keep the Heaviside function and the space
dependent function $G(\xi )$ in their lowest order, the anomalous dimension
is the same as the RG result. When the approximation of the Heaviside
function in its lowest order is removed, the value of the anomalous
dimension can be improved. It is expressed in eq.$\left( 91\right) $ and
Fig.1. }

\ 

{\normalsize \noindent }{\large {\bf Acknowledgments} }

\ 

{\normalsize G. Cheng is supported in part by the National Science
Foundation in China and by the NDSTPR. Foundation in China. The authors
would like to thank Miss J.L. Zhou, Dr. Zhang Yang and Dr. Han Lang for
their help in this paper. }

{\normalsize \qquad \newpage
}

\end{document}